\documentclass[a4paper]{jpconf}

\usepackage{amsfonts,amsmath,amssymb,bm,graphicx,subfigure,hyperref}

\begin{document}

\title{Neutrino spin oscillations in curved space-time under the influence of external fields}

\author{Maxim Dvornikov}

\address{Pushkov Institute of Terrestrial Magnetism, Ionosphere and Radiowave Propagation (IZMIRAN), 108840 Troitsk, Moscow, Russia}

\address{Physics Faculty, National Research Tomsk State University, 36 Lenin Avenue, 634050 Tomsk, Russia}

\ead{maxdvo@izmiran.ru}

\begin{abstract}
We study neutrino spin oscillations in background matter under the influence of strong electromagnetic and gravitational fields. The neutrino spin evolution is treated quasiclassically. We derive the effective Hamiltonian governing spin oscillations of a neutrino moving in the vicinity of a black hole and interacting with a relativistic magnetized accretion disk around a black hole. Applications for the studies of spin oscillations of astrophysical neutrinos are considered.
\end{abstract}

\section{Introduction}

Neutrinos play an important role in the evolution of various astrophysical objects like neutron stars, supernov\ae, quasars, etc. In these environments, neutrinos are subject to the interaction not only with a background matter and magnetic fields, but also with a strong gravitational field. The study of astrophysical neutrinos propagating in these backgrounds becomes important owing to the recent development of the multi-messenger approach in astronomy and astrophysics, when a correlation between a neutrino signal and, e.g., the detection of gravitational waves is searched.

Neutrinos are known to be mainly left-polarized particles in the standard model. However, the interaction with external fields can change their polarization and neutrinos become invisible for a terrestrial detector. This process is known as neutrino spin oscillations. Thus, the interaction of astrophysical neutrinos with external fields, which include gravity, can significantly modify the predictions of the multi-messenger approach.

In the paper, we review our results on the description of neutrino spin oscillations in non-trivial gravitational backgrounds under the influence of various external fields. In section~\ref{sec:GENFORM}, we derive the most general quasiclassical equation for the description of the neutrino spin evolution in these external fields. In sections~\ref{SM} and~\ref{sec:KERR}, we apply our results for the studies of neutrino spin oscillations in the vicinity of Schwarzschild and Kerr black holes (BH). Finally, in section~\ref{sec:CONCL}, we summarize the obtained results.

\section{Covariant description of the neutrino spin evolution\label{sec:GENFORM}}

In this section we shall construct the quasiclassical approach for
the description of the spin evolution of a Dirac neutrino moving in
a background matter under the influence of electromagnetic and gravitational
fields. In particular, we generalize the neutrino spin evolution equation
in matter and electromagnetic fields to include the effects of the nontrivial
geometry of space-time.

The equation describing the neutrino spin evolution in matter under the influence of an external electromagnetic field in Minkowski space-time was derived in~\cite{DvoStu02}. The most straightforward generalization of this equation to include a nontrivial gravitational background has the form~\cite{Dvo13},
\begin{equation}\label{eq:BMTcurvedst}
  \frac{\mathrm{D}S^{\mu}}{\mathrm{D}\tau} =
  2\mu
  \left(
    F^{\mu\nu}S_{\nu}-U^{\mu}U_{\nu}F^{\nu\lambda}S_{\lambda}
  \right) +
  \sqrt{2}G_{\mathrm{F}}E^{\mu\nu\lambda\rho}G_{\nu}U_{\lambda}S_{\rho}.
\end{equation}
where $S^{\mu}$ and $U^{\mu}$ are the four vectors of the spin and the velocity of a  neutrino, $F_{\mu\nu}$ is the skew-symmetric tensor of the electromagnetic field, $E^{\mu\nu\lambda\rho} = \varepsilon^{\mu\nu\lambda\rho}/\sqrt{-g}$ is the antisymmetric tensor is curved space-time, $g=\det(g_{\mu\nu})$ is the determinant of the metric tensor $g_{\mu\nu}$, $\mu$ is the neutrino magnetic moment, and $G_{\mathrm{F}} = 1.17\times 10^{-5}\,\text{GeV}^{-2}$ is the Fermi constant. The covariant derivative $\mathrm{D}/\mathrm{D}\tau$ along the world line in equation~\eqref{eq:BMTcurvedst} is taken with respect to the interval $\tau$. Equation~\eqref{eq:BMTcurvedst} should be completed by the geodesic equation for
the evolution of the four velocity, $\mathrm{D}U^{\mu}/\mathrm{D}\tau = 0$.

The neutrino interaction with a background matter is characterized by the four
vector $G^{\mu}$, which is a linear combination of hydrodynamic currents,
$J_{f}^{\mu}$, and polarizations, $\Lambda_{f}^{\mu}$, of background
fermions of the type $f$,
\begin{equation}\label{eq:Gmugen}
  G^{\mu} = \sum_{f}
  \left(
    q_{f}^{(1)}J_{f}^{\mu}+q_{f}^{(2)}\Lambda_{f}^{\mu}
  \right).
\end{equation}
We shall express $J_{f}^{\mu}$ using the invariant number density $n_{f}$,
i.e. the density in the rest frame of fermions, and the four velocity
$U_{f}^{\mu}$ of background fermions: $J_{f}^{\mu}=n_{f}U_{f}^{\mu}$.
The explicit expression of $\Lambda_{f}^{\mu}$ is given in~\cite{LobStu01}. The coefficients $q_{f}^{(1,2)}$ can be found in~\cite{DvoStu02}.

Then we introduce a locally Minkowskian frame by considering the vierbein $e_{\ \mu}^{a}$, $a=0,\dots,3$, satisfying the relations, $g_{\mu\nu} = e_{\ \mu}^{a}e_{\ \nu}^{b}\eta_{ab}$ and $\eta_{ab} = e_{a}^{\ \mu}e_{a}^{\ \nu}g_{\mu\nu}$, where $e_{a}^{\ \mu}$ is the inverse vierbein and $\eta_{ab}=\mathrm{diag}(1,-1,-1,-1)$ is the metric tensor in the Minkowski space-time. Decomposing all the quantities in equation~\eqref{eq:BMTcurvedst} in the vierbein basis, we get the following equation for the neutrino spin $s^{a}=e_{\ \mu}^{a}S^{\mu}$:
\begin{equation}\label{eq:BMTvierb}
  \frac{\mathrm{d}s^{a}}{\mathrm{d}t} =
  \frac{1}{\gamma}
  \left[
    G^{ab}s_{b}+2\mu
    \left(
      f^{ab}s_{b}-u^{a}u_{b}f^{bc}s_{c}
    \right) +
    \sqrt{2}G_{\mathrm{F}}\varepsilon^{abcd}g_{b}u_{c}s_{d}
  \right],
\end{equation}
where $G^{ab}=\eta^{ac}\eta^{bd}\gamma_{cde}u^{e}$ is the antisymmetric
tensor accounting for the gravitational interaction of neutrinos,
$\gamma_{abc}=\eta_{ad}e_{\ \mu;\nu}^{d}e_{b}^{\ \mu}e_{c}^{\ \nu}$
are the Ricci rotation coefficients (the semicolon
stays for the covariant derivative), $\gamma=U^{0}$, $g^{a}=e_{\ \mu}^{a}G^{\mu}=(g{}^{0},\mathbf{g})$
is the effective potential of the matter interaction in the vierbein
frame, $f_{ab}=e_{a}^{\ \mu}e_{b}^{\ \nu}F_{\mu\nu}=(\mathbf{e},\mathbf{b})$
is the electromagnetic field tensor in the vierbein frame, and $u^{a}=e_{\ \mu}^{a}U^{\mu}=(u^{0},\mathbf{u})$ is the neutrino velocity in the vierbein basis. We should also reformulate the dynamics of $U^{\mu}$ in the vierbein frame.
One can obtain the following evolution equation for $u^{a}$:
\begin{equation}\label{eq:Uvierb}
  \frac{\mathrm{d}u^{a}}{\mathrm{d}t} = \frac{1}{\gamma}G^{ab}u_{b}.
\end{equation}
The details of the derivation of equations~(\ref{eq:BMTvierb}) and~(\ref{eq:Uvierb})
can be found in~\cite{Dvo13,Dvo06}.

The three vector of the neutrino polarization $\bm{\zeta}$ is defined as
\begin{equation}\label{eq:nuspinzeta}
  s^{a} =
  \left(
    (\bm{\zeta}\cdot\mathbf{u}),
    \bm{\zeta}+\frac{\mathbf{u}(\bm{\zeta}\cdot\mathbf{u})}{1+u^{0}}
  \right).
\end{equation}
Using
equations~(\ref{eq:BMTvierb}) and~(\ref{eq:nuspinzeta}) one finds the
evolution equation for $\bm{\zeta}$ as
\begin{equation}\label{eq:nuspinrot}
  \frac{\mathrm{d}\bm{\zeta}}{\mathrm{d}t} = \frac{2}{\gamma}[\bm{\zeta}\times\mathbf{G}],
\end{equation}
where
\begin{equation}\label{eq:vectG}
  \mathbf{G} = 
  \frac{1}{2}
  \left[
    \mathbf{b}_{g}+\frac{1}{1+u^{0}}
    \left(
      \mathbf{e}_{g}\times\mathbf{u}
    \right)
  \right] +
  \frac{G_{\mathrm{F}}}{\sqrt{2}}
  \left[
    \mathbf{u}
    \left(
      g^{0}-\frac{(\mathbf{g}\mathbf{u})}{1+u^{0}}
    \right) -
    \mathbf{g}
  \right] +
  \mu
  \left[
    u^{0}\mathbf{b}-\frac{\mathbf{u}(\mathbf{u}\mathbf{b})}{1+u^{0}}+(\mathbf{e}\times\mathbf{u})
  \right].
\end{equation}
Here $\mathbf{e}_{g}$ and $\mathbf{b}_{g}$ are the components of
the tensor $G_{ab}$: $G_{ab}=(\mathbf{e}_{g},\mathbf{b}_{g})$. To
derive equation~(\ref{eq:vectG}) we use the fact that $u_{a}u^{a}=1$.

\section{Neutrino spin oscillations in Schwarzschild metric\label{SM}}

Let us apply general equations~\eqref{eq:nuspinrot} and~\eqref{eq:vectG} to describe the neutrino spin evolution in the vicinity of a non-rotating BH. In this section, we shall omit the influence of the background matter and the electromagnetic field on neutrino oscillations.

When we study the gravitational field of a non-rotating BH, the interval is known to be expressed with help of the
Schwarzschild metric,
\begin{equation}\label{schwarz}
  d\tau^2=A^{2}dt^2-A^{-2}dr^2-
  r^2(d\theta^2+\sin^2\theta d\phi^2),
\end{equation}
where $A=\sqrt{1-r_g/r}$, $r_g = 2M$ is the Schwarzschild radius, and $M$ is the BH mass. In Eq.~\eqref{schwarz} we use the spherical coordinates $(r,\theta,\phi)$.

The neutrino spin precession is determined by the vector
$\bm{\Omega}=\mathbf{G}/\gamma$. The components of this vector can
be found on the basis of equation~\eqref{eq:vectG} (see also~\cite{Dvo06}),
\begin{equation}\label{Omega1}
  \Omega_1 = \frac{1}{2}v_\phi\cos\theta,
  \quad
  \Omega_2 =v_\phi\sin\theta\frac{1}{2}
  \left(
    -A+\frac{\gamma}{(1+\gamma A)}
    \frac{r_g}{2r}
  \right),
  \quad
  \Omega_3 =v_\theta\frac{1}{2}
  \left(
    A-\frac{\gamma}{(1+\gamma A)}
    \frac{r_g}{2r}
  \right),
\end{equation}
where $\mathbf{v}=(v_r,v_\theta,v_\phi)$ are the components of the
world velocity.

Let us discuss a neutrino orbiting BH. For simplicity we
consider only circular orbits with the radius $R$. We may restrict
ourselves to the consideration of the orbits lying only in the
equatorial plane ($\theta=\pi/2$ or equivalently $v_\theta=0$)
because of the spherical symmetry of the gravitational field. In
this case $\Omega_1 = \Omega_3 = 0$ in equation~\eqref{Omega1}. The expressions for the neutrino angular velocity
and $\gamma$ have the form,
\begin{equation}\label{vphico}
  v_\phi=\frac{d\phi}{dt}=\sqrt{\frac{r_g}{2R^3}},
  \quad
  \gamma^{-1}=\frac{d\tau}{dt}=\sqrt{1-\frac{3r_g}{2R}}.
\end{equation}
It should be noted that the vierbein four velocity now takes the
form, $u^a=(\gamma A, 0, 0, \gamma v_\phi r)$. One can verify that
$u^a u_a=1$ with help of equation~\eqref{vphico}. We also mention that equation~\eqref{eq:Uvierb} is also identically satisfied since $G^{ab}u_{b}=0$. Therefore a neutrino has a constant four velocity with respect to the vierbein frame.

Using equation~\eqref{Omega1}, we can
rewrite the remaining nonzero component of $\bm{\Omega}$ in the more simple form, $\Omega_2=-v_\phi/2\gamma$. We suppose that initially a neutrino is left-handed, i.e. its
initial spin vector is antiparallel to the particle's velocity.
According to equations~\eqref{eq:nuspinrot} and \eqref{Omega1},
the neutrino spin rotates around the second axis. Therefore we can derive the effective Schr\"odinger equation for neutrino spin oscillations in the gravitational field of a
non-rotating BH,
\begin{equation}\label{HeffBH}
  \mathrm{i}\frac{\mathrm{d}\nu}{\mathrm{d}t} = H_{\mathrm{eff}}\nu,
  \quad
  H_\mathrm{eff}= -(\bm{\sigma}\cdot\bm{\Omega}) =
  \begin{pmatrix}
    0 & -\mathrm{i}\Omega_2 \\
    \mathrm{i}\Omega_2 & 0
  \end{pmatrix},
\end{equation}
where $\nu^{\mathrm{T}}=(\nu_\mathrm{R},\nu_\mathrm{L})$ is the effective neutrino wave function, $H_{\mathrm{eff}}$ is the effective Hamiltonian, and $\bm{\sigma} = (\sigma_1,\sigma_2,\sigma_3)$ are the Pauli matrices.

\begin{figure}
  \centering
  \includegraphics[scale=.2]{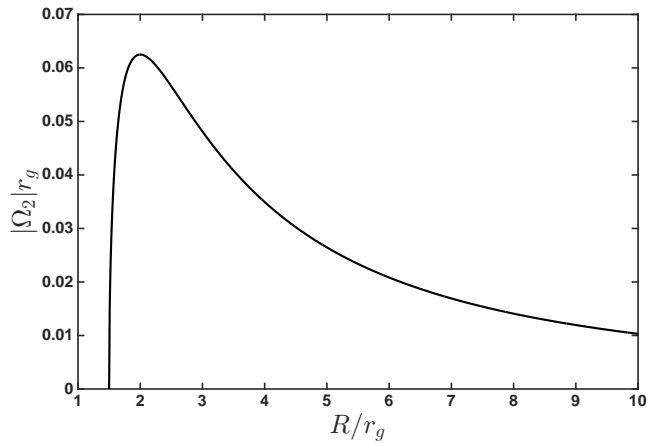}
  \protect
  \caption{
    Neutrino spin oscillations frequency versus the radius
  of the neutrino orbit.\label{freq} 
  }
\end{figure}

Using equation~\eqref{HeffBH}, we obtain the expression for the transition probability for neutrino spin oscillations, $P(t)=\sin^2(\Omega_2 t)$. One can see that there is a full mixing in our case, and thus, the neutrino transition probability can achieve a unit value. Let us plot the frequency of neutrino spin oscillations
versus the radius of the orbit. It is possible to see in
figure~\ref{freq} that $|\Omega_2|=0$ at $R=1.5 r_g$ and
$|\Omega_2|\to 0$ at $R\to\infty$. One can also conclude that $|\Omega_2|$ has its maximal value, which is equal to $6.25\times 10^{-2} r_g^{-1}$, at $R=2 r_g$. Let us evaluate the number of revolutions $N$, that a neutrino should make, necessary for the total spin
flip. Using equation~\eqref{vphico} we can find that $N = \gamma =2$ at $R=2 r_g$. It is interesting to evaluate the characteristic period of the
neutrino spin oscillations, i.e. the oscillations length. For $M=10
M_\odot$ at $R=2 r_g$ we get for $T=\pi/|\Omega_2|\approx
4.94\times 10^{-3}\thinspace\text{s}$.

\section{Neutrino propagation in the vicinity of a rotating black hole\label{sec:KERR}}

Now let us describe neutrino spin oscillations
in matter and a magnetic field when a particle moves in the vicinity
of a rotating BH. As in section~\ref{SM}, we shall consider circular neutrino motion and derive the effective Schr\"{o}dinger equation as well as find the transition probability for spin oscillations.

The space-time of a rotating BH can be described by the Kerr metric.
In Boyer-Lindquist coordinates $x^\mu = (t,r,\theta,\phi)$, this metric has the form,
\begin{equation}\label{eq:Kerrmetr}
  \mathrm{d}\tau^{2} =
  \left(
    1-\frac{rr_{g}}{\Sigma}
  \right)
  \mathrm{d}t^{2} +
  2\frac{rr_{g}a\sin^{2}\theta}{\Sigma}\mathrm{d}t\mathrm{d}\phi -
  \frac{\Sigma}{\Delta}\mathrm{d}r^{2} -
  \Sigma\mathrm{d}\theta^{2} -
  \frac{\Xi}{\Sigma}\sin^{2}\theta\mathrm{d}\phi^{2},
\end{equation}
where
\begin{equation}\label{eq:metrpar}
  \Delta=r^{2}-rr_{g}+a^{2},
  \quad
  \Sigma=r^{2}+a^{2}\cos^{2}\theta,
  \quad
  \Xi=\left(r^{2}+a^{2}\right)\Sigma+rr_{g}a^{2}\sin^{2}\theta.
\end{equation}
The angular momentum of BH is $J=Ma$.
Coordinates $x^\mu$ in equations~\eqref{eq:Kerrmetr} and~\eqref{eq:metrpar} correspond to the world frame rather than to the vierbein frame.

We shall discuss BH surrounded by matter forming an accretion
disk in equatorial plane. In this case only $U_{f}^{0}$ and $U_{f}^{\phi}$ are nonzero.
Moreover we shall consider a stationary accretion. By the symmetry
reasons, the quantities $n_{f}$, $U_{f}^{0}$, and $U_{f}^{\phi}$
can be functions of $r$ and $\theta$ only. Assuming that the temperature of background fermions is sufficiently high, we can take that the background matter is unpolarized. In this case, the effective neutrino interaction with matter is characterized by the four vector,
\begin{equation}\label{eq:gKerr}
  g^{a} =
  n_{\mathrm{eff}}
  \left(
    rU_{f}^{0}\sqrt{\frac{\Delta}{\Xi_{0}}},0,0,
    \frac{U_{f}^{\phi}\Xi_{0}-arr_{g}U_{f}^{0}}{r\sqrt{\Xi_{0}}}
  \right),
  \quad
  n_{\mathrm{eff}} = \sum_{f}q_{f}^{(1)}n_{f},
\end{equation}
where $\Xi_{0}=\Xi(\theta=\pi/2)=r^{4}+a^{2}r(r+r_{g})$.
In equation~(\ref{eq:gKerr}) we assume that all types of background fermions have the same four velocity.

Now let us specify the structure of the electromagnetic field. We
suggest that, at relatively great distance from BH, there is a constant
and uniform magnetic field parallel to the rotation axis of BH and
having the strength $B_{0}$. This magnetic field can be created by the
plasma motion in the accretion disk. The structure of the axially symmetric
electromagnetic field in a curved space-time having such an asymptotics
was studied in~\cite{Wal74}, where one can find the nonzero components of the
vector potential.
%
On the basis of the explicit form of the vector potential (see, e.g.,~\cite{Dvo13}), the electric and magnetic field strengths can be calculated as $F_{\mu\nu}=\partial_{\mu}A_{\nu}-\partial_{\nu}A_{\mu}$.

As in the case of Schwarzschild metric, studied in section~\ref{SM}, here one can show that the four velocity in the vierbein frame $u^a$ is constant if we study circular neutrino orbits. Using this fact, we can reformulate the neutrino
spin dynamics using the effective Schr\"{o}dinger equation~\eqref{HeffBH}. The vector $\bm{\Omega}$, which defines the neutrino spin precession, has the following nonzero components:
\begin{align}
  \Omega_{2} & =
  \frac{1}{2\gamma r_{g}}
  \left\{
    \mp\frac{1}{\sqrt{2}x^{3/2}} -
    \mu B_{0}r_{g}
    \frac{2\sqrt{2}x^{2}(x-1)\pm\alpha\sqrt{x}(2x-1)+\sqrt{2}\alpha^{2}}
    {x^{3/2}\sqrt{2x^{3}-3x^{2}\pm2\sqrt{2}\alpha x^{3/2}}}
  \right\},
  \nonumber \\
  \Omega_{3} & =
  \frac{G_{\mathrm{F}}n_{\mathrm{eff}}}{\sqrt{2}\gamma}
  \frac{\pm U_{f}^{0}-r_{g}U_{f}^{\phi}(\sqrt{2}x^{3/2}\pm\alpha)}
  {\sqrt{2x^{3}-3x^{2}\pm2\sqrt{2}\alpha x^{3/2}}},
  \label{eq:Omega23}
\end{align}
where $x=r/r_{g}$ and $\alpha=a/r_{g}$. The upper sign refers to
the direct orbits, i.e. when a neutrino corotates with BH, and the
lower sign corresponds to retrograde ones (a particle counter-rotates).

Supposing that $B_{0}$ and $n_{\mathrm{eff}}$ do not depend on time,
we can solve the Schr\"{o}dinger equation analytically. Assuming that
initially only left-polarized neutrinos are present, i.e. $\nu^{\mathrm{T}}(0)=(0,1)$,
we get the probability to detect a right-polarized neutrino in the
form,
\begin{equation}\label{eq:Ptr}
  P(t) = P_{\mathrm{max}}\sin^{2}
  \left(
    \frac{\pi}{L_{\mathrm{osc}}}t
  \right),
  \quad
  P_{\mathrm{max}} = \frac{\Omega_{2}^{2}}{\Omega_{2}^{2}+\Omega_{3}^{2}},
  \quad
  L_{\mathrm{osc}} = \frac{\pi}{\sqrt{\Omega_{2}^{2}+\Omega_{3}^{2}}}.
\end{equation}
Here $P_{\mathrm{max}}$ is the amplitude and $L_{\mathrm{osc}}$ is the length of oscillations .

\begin{figure}
  \centering
  \includegraphics[scale=0.25]{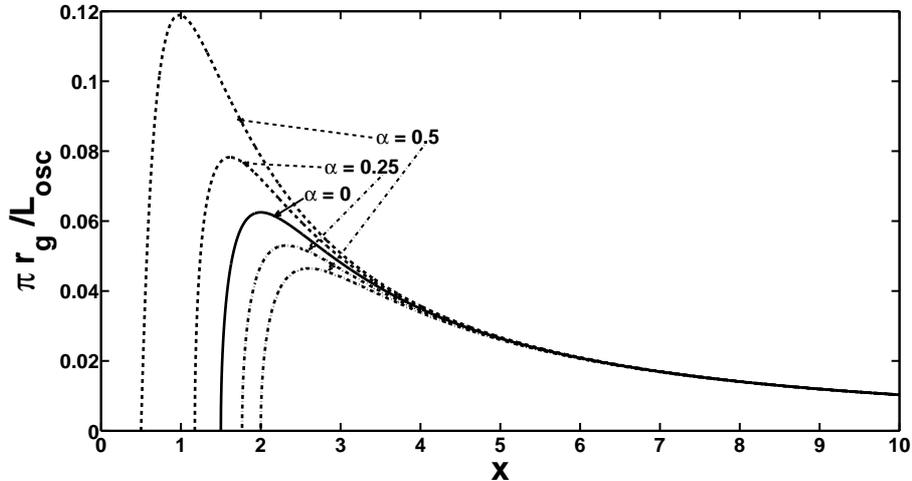}
  \caption{
  The function $\pi r_{g}/L_{\mathrm{osc}}=|\Omega_{2}|r_{g}$ versus
  $x=r/r_{g}$ for $\alpha=0$, $0.25$, and $0.5$. The case $\alpha=0$
  (solid line) corresponds to a nonrotating BH, described by the Schwarzschild
  metric studied in section~\ref{SM}. For $\alpha\neq0$,
  dashed lines are depicted for direct orbits and dash-dotted lines
  for retrograde ones.}
  \label{fig:Omega2vsx}
\end{figure}

Considering the situation when $n_\mathrm{eff} = 0$ and $B_0 = 0$, i.e. only the gravitational interaction is accounted for, we present the dependence $\pi r_{g}/L_{\mathrm{osc}}=|\Omega_{2}|r_{g}$
versus $x=r/r_{g}$ for different values of $\alpha$ in figure~\ref{fig:Omega2vsx}. It can be seen
that for retrograde orbits the maximal value of $|\Omega_{2}|$ decreases
with $\alpha$, whereas for direct orbits this value increases with
$\alpha$.

If we discuss the maximal possible angular momentum of BH corresponding
to $\alpha=1/2$, we get from figure~\ref{fig:Omega2vsx} that for
a direct neutrino orbit with $r\approx r_{g}$, the frequency of neutrino
spin oscillations reaches its maximal value $\Omega_{2}^{+}\approx0.12r_{g}^{-1}$.
For a retrograde orbit with $r\approx2.60r_{g}$ we get the maximal
frequency $\Omega_{2}^{-}\approx0.05r_{g}^{-1}$. Considering BH with
$M=10M_{\odot}$, we obtain the corresponding oscillations lengths,
$L_{\mathrm{osc}}^{+}\approx7.85\times10^{7}\thinspace\text{cm}$
and $L_{\mathrm{osc}}^{-}\approx1.88\times10^{8}\thinspace\text{cm}$. On the basis of these oscillations lengths one gets the times of the neutrino spin-flip, $T_\mathrm{sf} = L_{\mathrm{osc}}/2$, as $T_\mathrm{sf}^{+} \approx 1.31 \times 10^{-3}\thinspace\text{s}$
and $T_\mathrm{sf}^{-} \approx 3.13 \times 10^{-3}\thinspace\text{s}$.

Note that for $\alpha=1/2$ the minimal possible radius of a stable
retrograde orbit is equal to $r_{\mathrm{min}}=4.5r_{g}$.
Therefore the obtained $\Omega_{2}^{-}$ corresponds to an unstable
orbit. In case of a direct orbit $r_{\mathrm{min}}\to r_{g}/2$. In
figure~\ref{fig:Omega2vsx} we get that in this case $\Omega_{2}=0$.

\section{Conclusion\label{sec:CONCL}}

In this work, we developed the quasiclassical approach for the description of neutrino spin oscillations in background matter under the influence of electromagnetic and gravitational fields. As applications of the obtained results, we considered spin evolution of neutrinos orbiting Schwarzschild and Kerr BH. We also discussed the possibility of the neutrino interaction with an ultrarelativistic magnetized accretion disk around BH.

Recently, the dynamics of the fermion spin in gravitational fields was studied in~\cite{ObuSilTer17}. In that work, the spin evolution was based on the analysis of the Dirac equation in curved space-time. Although the results of~\cite{ObuSilTer17} are analogous to that of our works~\cite{Dvo13,Dvo06}, where we used the quasiclassical approximation, we can describe the neutrino spin evolution in an arbitrary gravitational field without referring to the Dirac equation. This fact makes the approach in~\cite{Dvo13,Dvo06} to be more attractive for possible applications.

I am thankful to organizers of DSPIN-17 for the invitation and a financial support,
as well as to the Tomsk State University Competitiveness Improvement
Program and RFBR (research project No.~15-02-00293) for a partial support.

\end{document}